\documentclass[5p,a4paper,10pt]{elsarticle}

\usepackage{graphicx}
\usepackage{amsmath, amsthm, amssymb}
\usepackage[utf8]{inputenc}
\usepackage{natbib}
\usepackage{stfloats}
\usepackage{hyperref}
\hypersetup{
    bookmarks=true,         % show bookmarks bar?
    pdffitwindow=false,     % window fit to page when opened
    pdfstartview={FitH},    % fits the width of the page to the window
    pdftitle={Crucible-free Pulling of Germanium Crystals},    % title
    pdfauthor={Michael Wünscher, Anke Lüdge, Helge Riemann},     % author
    pdfcreator={Michael Wünscher},   % creator of the document
    pdfproducer={Michael Wünscher}, % producer of the document
    pdfkeywords={Heat transfer} {Floating zone technique} {Growth from melt} {Semiconducting germanium}
    pdfnewwindow=true,      % links in new window
    colorlinks=true,       % false: boxed links; true: colored links
    linkcolor=red,          % color of internal links
    citecolor=green,        % color of links to bibliography
    filecolor=magenta,      % color of file links
    urlcolor=cyan           % color of external links
}

\begin{document}
%opening
\begin{frontmatter}
\title{Crucible-free Pulling of Germanium Crystals}

\author{Michael Wünscher}
\ead{wuenscher@ikz-berlin.de}
\author{Anke Lüdge}
\author{Helge Riemann}

\address{Leibniz Institute for Crystal Growth, Max-Born-Str. 2, 12489 Berlin, Germany\vspace{-1cm}}

\begin{abstract}
Commonly, germanium crystals are grown after the Czochralski (CZ) method. The crucible-free pedestal and floating zone (FZ) methods, which are widely used for silicon growth, are hardly known to be investigated for germanium. 
The germanium melt is more than twice as dense as liquid silicon, which could destabilize a floating zone. Additionally, the lower melting point and the related lower radiative heat loss is shown to reduce the stability especially of the FZ process with the consequence of a screw-like crystal growth. 
We found that the lower heat radiation of Ge can be compensated by the increased convective cooling of a helium atmosphere instead of the argon ambient. Under these conditions, the screw-like growth could be avoided.  Unfortunately, the helium cooling deteriorates the melting behavior of the feed rod. Spikes appear along the open melt front, which touch on the induction coil. In order to improve the melting behavior, we used a lamp as a second energy source as well as a mixture of Ar and He. With this, we found a final solution for growing stable crystals from germanium by using both gases in different parts of the furnace.
The experimental work is accompanied by the simulation of the stationary temperature field. The commercially available software FEMAG-FZ is used for axisymmetric calculations. Another tool for process development is the lateral photo-voltage scanning (LPS), which can determine the shape of the solid-liquid phase boundary by analyzing the growth striations in a lateral cut of a grown crystal. In addition to improvements of the process, these measurements can be compared with the calculated results and, hence, conduce to validate the calculation.
\end{abstract}

\end{frontmatter}

\section{Introduction}
\footnotetext{link to the original document: \href{http://dx.doi.org/10.1016/j.jcrysgro.2010.10.200}{10.1016/j.jcrysgro.2010.10.200}}
Germanium crystals are mainly used for infrared optics, gamma/X-ray detectors or as substrate for high-power solar cells. Nevertheless the market is much smaller, than it is for silicon, due to the rareness of germanium and the associated high price. These crystals are grown by the CZ technique. Especially for high-purity crystals as used in detector technologies with a net doping of $10^{10}$ atoms per $cm^3$, a sophisticated CZ process is needed to reach this goal. The FZ  and pedestal techniques are known to reduce impurities depending on the segregation coefficient. Additionally contamination with impurities from the machine is reduced, because only the gas is in direct contact with the melt. The goal of our research is to find a way to use the advantages of the crucible-free methods for germanium.

\begin{figure}[hbtp]
\centering{
\includegraphics[width=7cm]{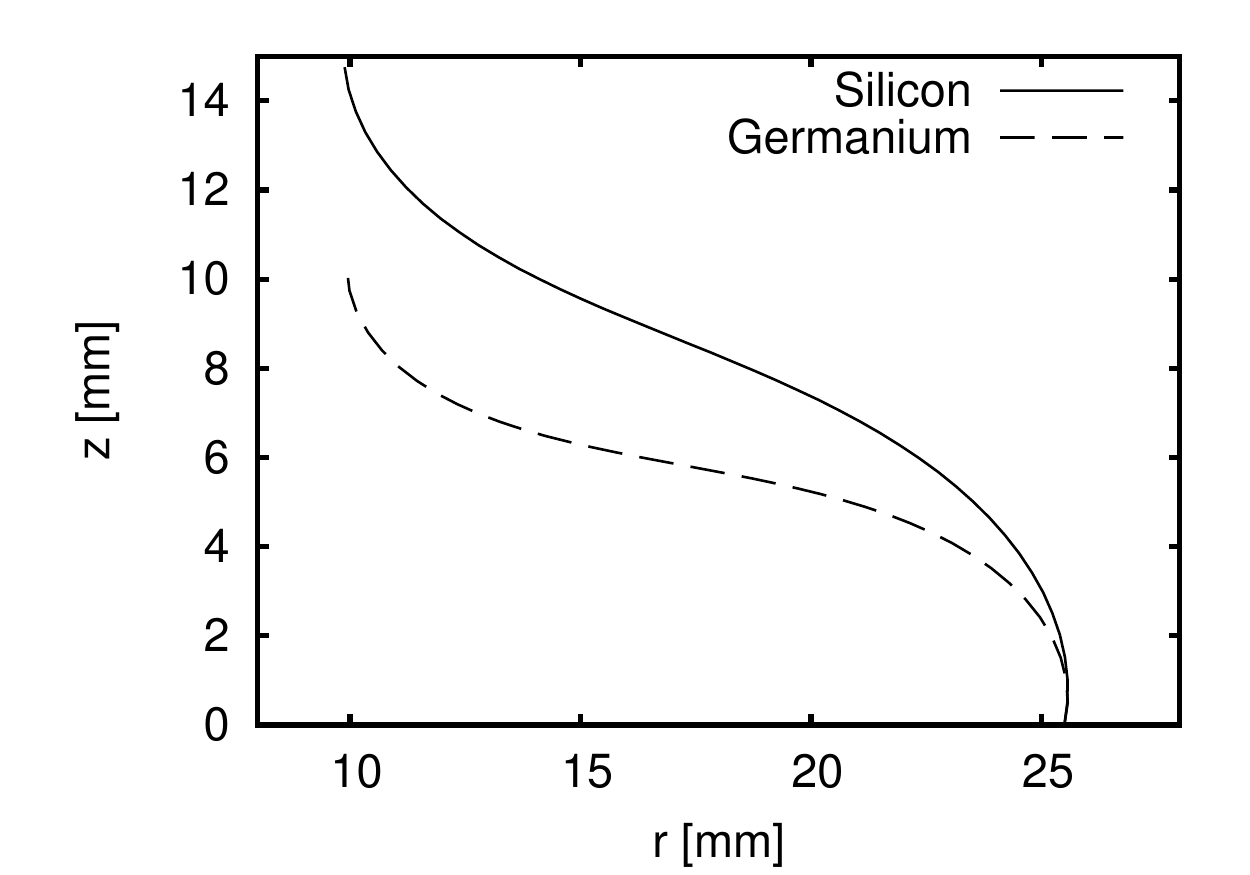}
}
\caption{Comparison of the melt surface between germanium and silicon based on calculation of the Laplace-Young equation \cite{Luedge2010}\label{fig:surfacecomparison}}
\end{figure}

From the viewpoint of growing germanium crystals with the FZ technique, the main disadvantage of Ge compared to silicon is the double melt density for a nearly equal surface tension. This reduces the height of the melt zone compared to silicon, which can be seen from the numerical simulation shown in figure \ref{fig:surfacecomparison}. Investigations of the FZ process show that the lower melting point increases the risk of thermo-technical destabilization during the growth, which yields to screw-like growth of the crystal. The heat loss at melting point temperature is four times smaller for germanium compared to silicon. To overcome the screw-like growth we had to optimize our silicon FZ process for the needs of germanium.

\begin{figure*}[tbp]
\centering{
\includegraphics[width=15cm]{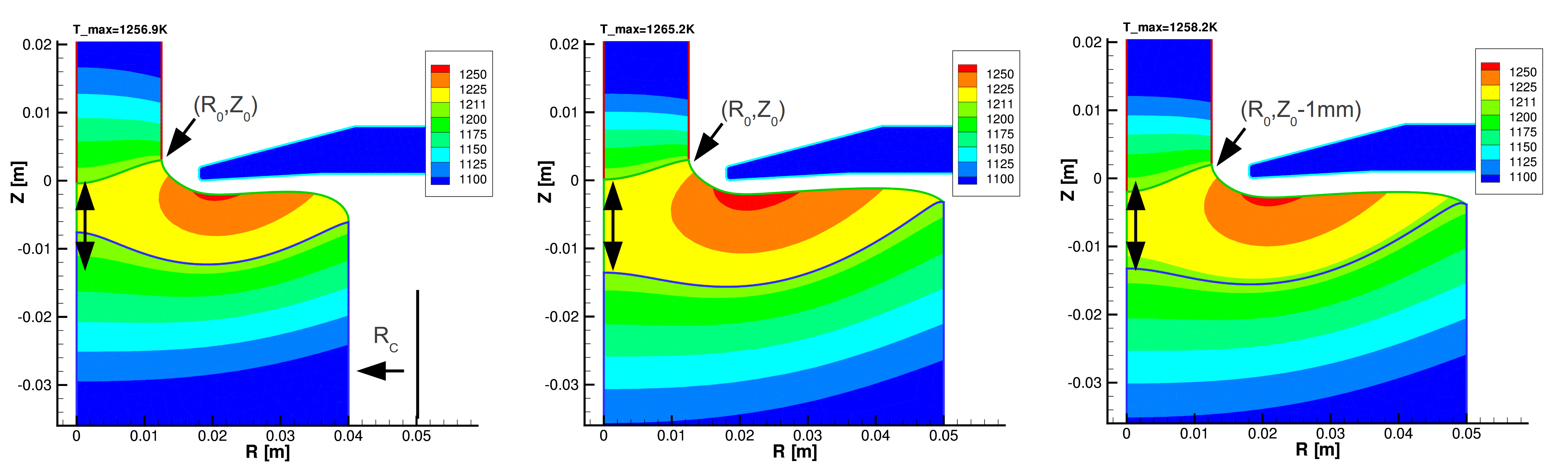}
}
\caption{Calculated temperature field for a 25mm crystal; left: smaller feed rod; middle: optimized starting configuration; right:  triple point moved downward by 1mm\label{fig:tempfield}}
\end{figure*}

The instability did not appear, when using the pedestal technique. Here, the challenge was to find a process which could be run without ``freezing''  in the center. For that purpose the software FEMAG-FZ was adapted to calculate the pedestal process. The results were compared with the experiments to validate the model.

\section{Pedestal}
\begin{figure}[thbp]
\centering{
\includegraphics[width=5cm]{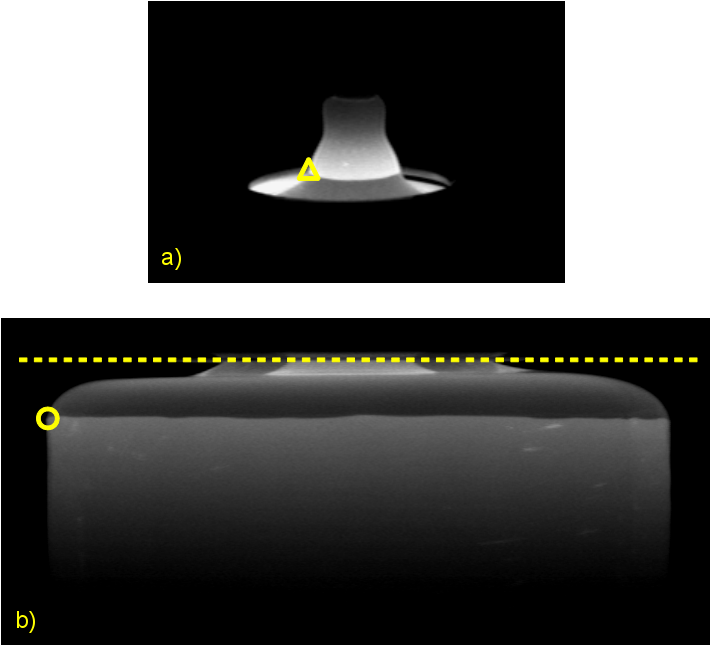}
}
\caption{Picture of the pedestal process for a 15mm crystal pulled from a 50mm feed rod. (a) Top view with upper triple point (triangle); (b) view from below the induction coil (dashed line shows bottom edge) with the marked lower triple point (circle)\label{fig:prozessbild}}
\end{figure}

The pedestal method is more stable than the FZ process.
Clearly, the diameter of the growing crystal can never be bigger than the diameter of the hole in the induction coil. In figure  \ref{fig:prozessbild}a) the crystal is pulled upwards from the melt through the induction coil, which melts the feed rod from the top (see fig. \ref{fig:prozessbild}b ).

We have examined the conditions for growing germanium crystals from bigger feed rods (diameter $>$ 50mm) with the pedestal method. For this method we have to find the way between two process boundaries, spilling out of the melt or the freezing of the melt in the central region. The freezing of the center is hard to predict, because it is covered with melt. Numerical studies can help to optimize the process and to get a better physical understanding of the process. It can also qualitatively help to change the process parameters.

In figure \ref{fig:tempfield} three different realizations of a growth process are shown. A reduction in the feed rod diameter
(compare left with middle picture) has two negative effects. On the one hand, it reduces the distance between the two phase boundaries, which increases the risk of freezing the melt in the center and consequently break the crystal growth process. On the other hand, it increases the melt meniscus, which increases the risk of spillage. Through lowering the crystal (cmp. middle with right), there is a tendency of the temperature field to shorten the meltable volume. The maximum temperature of the melt is lowered as well. From this solution where the melting point temperature is not reached along the melting front, one can see that freezing is likely possible. Furthermore, it could be supposed that the feed rod would not be fully molten during the experiment. Such process conditions were observed in some experiments. From this we can summarize that an experiment should be conduct with a more stretched zone to get more power into the central part.

In order to check the calculation results, one needs experimental data. The lateral photo-voltage scanning (LPS) method allows to measure the striations of the crystal. These striations follow the solidification interface, which nearly equates with the melting point isotherm of the calculated temperature field.

\begin{figure}[bhtp]
\centering{
\includegraphics[width=9cm]{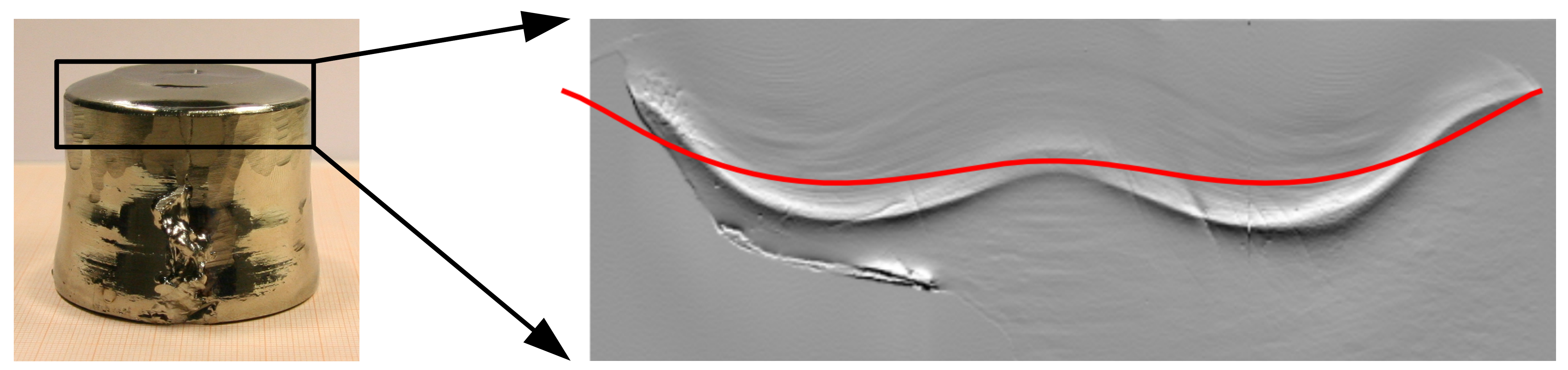}
}
\caption{Left: feed rod 50mm Right: LPS measurement of the striations with calculated phase boundary (solid line)\label{fig:laengsschnitt}}
\end{figure}

\begin{figure}[hbtp]
\centering{
\includegraphics[width=5cm]{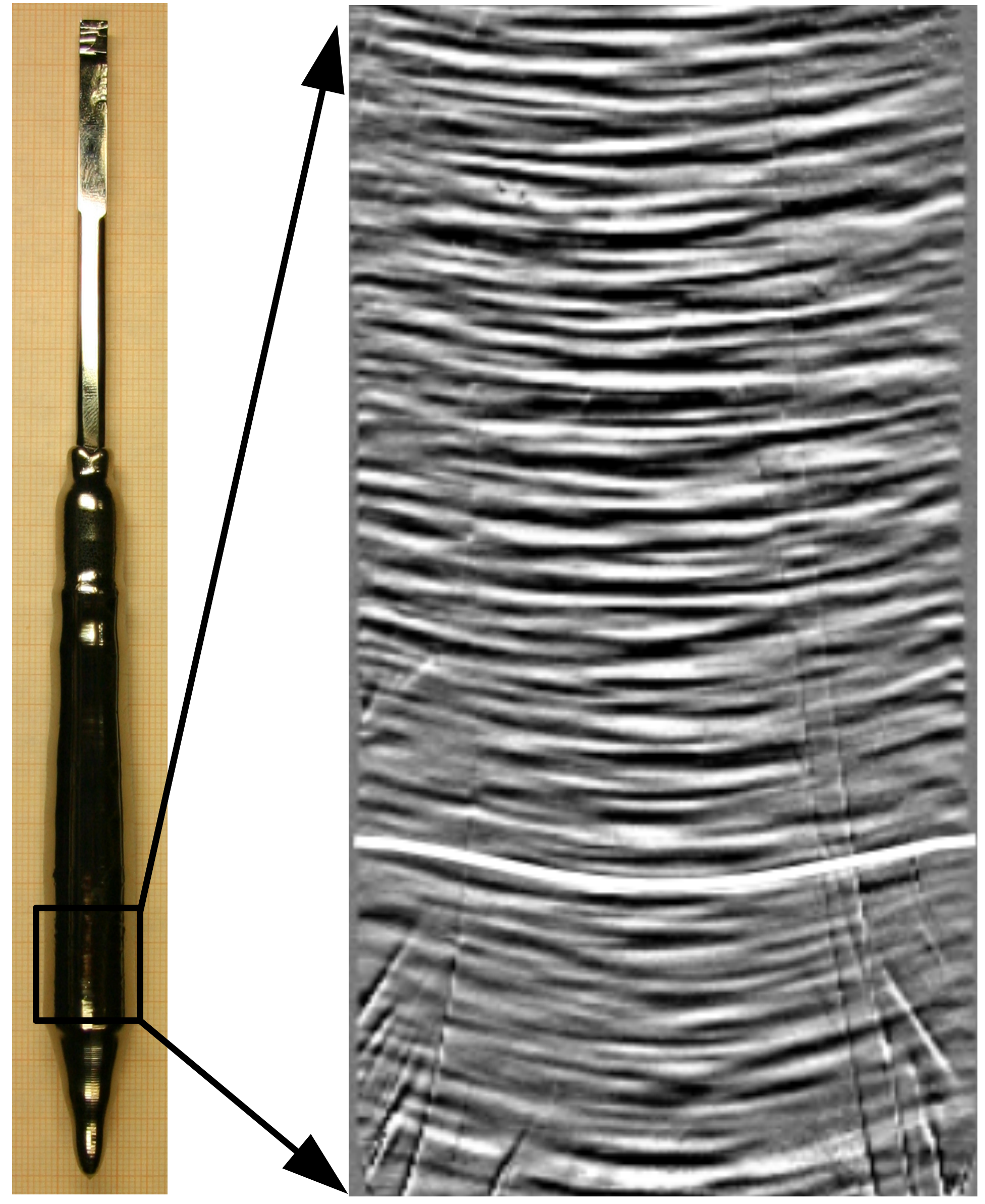}
}
\caption{Left: crystal with 15mm diameter Right: LPS measurement of striations with calculated phase boundary (white)\label{fig:laengsschnittkristall}}
\end{figure}

As input parameters from the experiment, the simulation takes the position of the triple points and the melt surface as a 2D curve, which are measured from process pictures like figure \ref{fig:prozessbild}. The process picture is not symmetric, but, using the left or  right melt surface does not show big influence on the simulation results. The melting point isotherms can now be put into the striation pictures. For the crystal (fig. \ref{fig:laengsschnittkristall}), there is a very good agreement between experiment and calculation. The agreement for the feed rod is not as good (fig. \ref{fig:laengsschnitt}), but it shows qualitatively the ``w'' shape from the experiment. Also the hill in the center has the same height, therefore, freezing can be well predicted. There are two main deviations from the experiment. The end of the isotherm is more steep than in experiment and the minimum of the isotherm is not big enough. Maybe both deviations are related to the unconsidered melt convection.

\vspace{-0.7cm}
\section{Floating Zone}
\vspace{-0.3cm}
\begin{figure}[hbtp]
\centering{
\includegraphics[width=7cm]{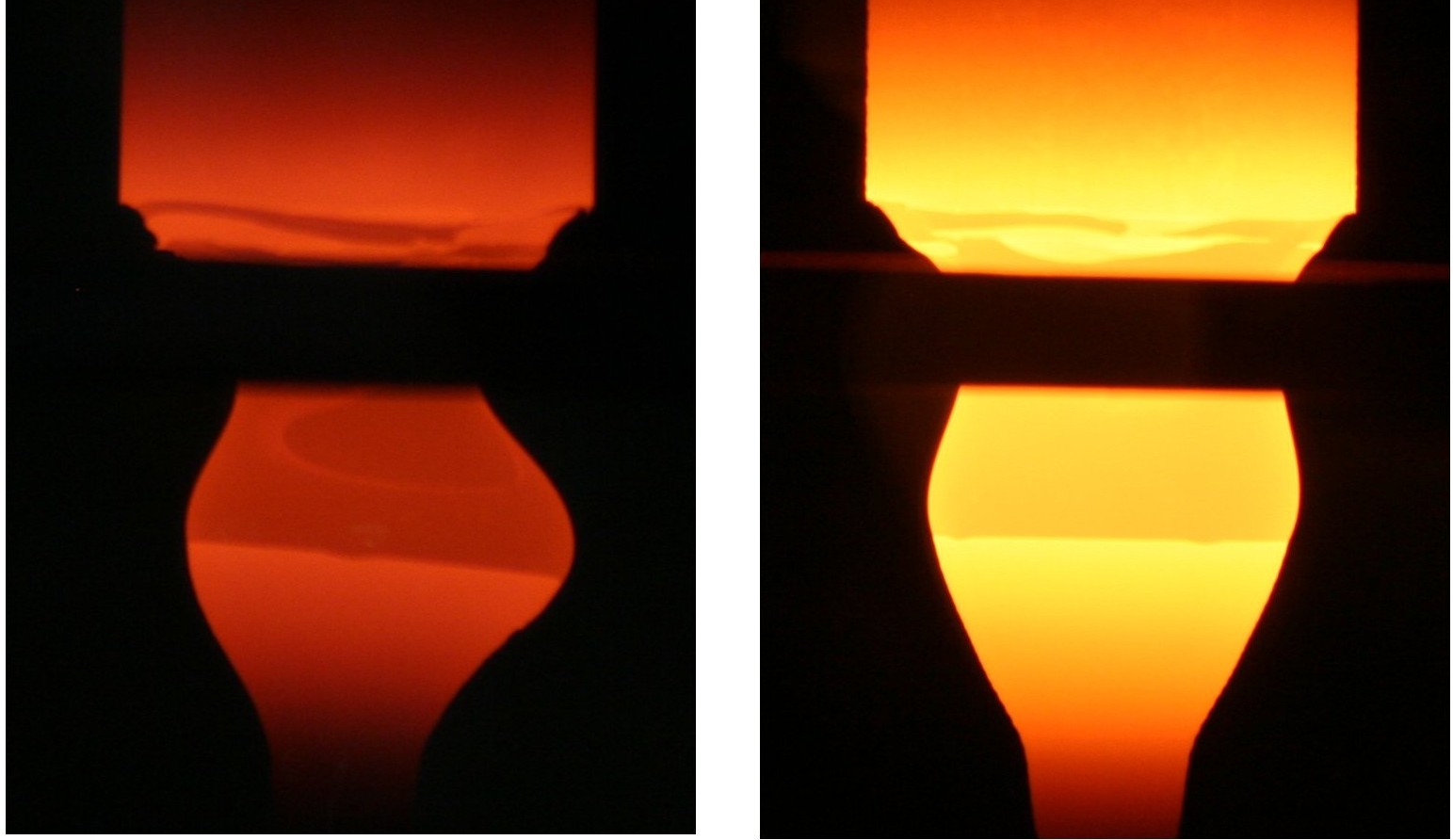}
}
\caption{Comparison between germanium (left) and silicon (right) for the same experimental configuration (feed rod diameter 20mm) at the beginning of the process. The silicon crystal was grown for several centimeters, whereas the germanium process spilled out shortly after.\label{fig:prozesssige}}
\end{figure}
The crucible-free floating zone method needs a thermally and mechanically stable melt zone. Therefore, a smooth melting of the feed rod as well as a tunable shape of melt zone and crystallization phase boundary during the whole process is needed.

As starting point for the growth of germanium crystals, we used a well-performing setup from silicon crystal growth. For silicon, it is easily possible to grow crystals of 20mm and more with this inductor. For germanium, we got a destabilization of the melt zone during the growth of the cone. Both are compared in figure \ref{fig:prozesssige}.  The phase boundary line tilts a bit and a buckle is growing out (fig. \ref{fig:prozesssige} left). After a few millimeters of pulling, the melt spills out at this buckle. It was only possible to grow small crystal, less than the hole diameter with this induction coil. In the next experiment we tried to tune the accessible parameter during the experiment, e.g. rotation rate and  pulling speed. No stabilization was achieved with this approach.

\begin{figure}[b!tph]
\centering{
\includegraphics[width=7cm]{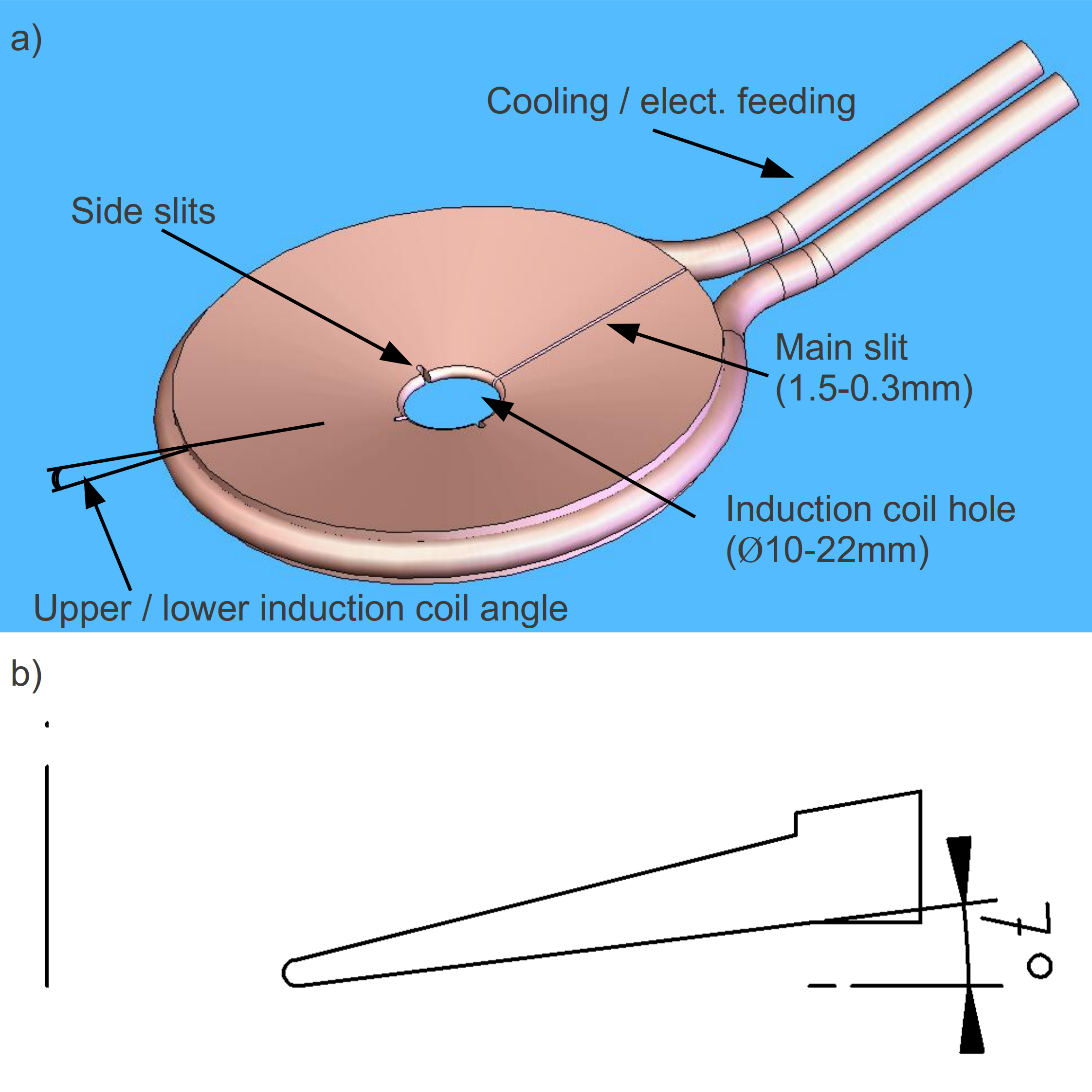}
}
\caption{a) Parameter for the induction coil design b) Induction coil with a conical angle of $7^\circ$\label{fig:induktor}}
\end{figure}

In a next step a compensation of the differences in material properties between silicon and germanium was tried. Therefore, the  silicon needle-eye process was optimized using the dependencies of numerical simulations with germanium material parameter. As a result new coil designs were tried. The important parameters of the induction coil are shown in figure \ref{fig:induktor}a.

The highest inductive heating is achieved below the edge of the induction coil. Therefore, geometrical changes in this region are very effective. At first the hole diameter of the coil was reduced from 15mm to 10mm. In principle the coil was scaled down to fit the reduced melt height. The modification shows the same results of screw-like growth starting from a diameter of 15mm. With the reduction of the main slit and the additional side slits the rotational symmetry of the coil was increased with no effect.

A new induction coil with a conical bottom side of $7^\circ$ angle (fig. \ref{fig:induktor}b), formerly flat, was built. Together with an increased hole diameter of 17.5mm, germanium crystals of 21mm were pulled. A further increase in diameter caused the occurrence of the described instable growth.

The last design came from considering the difference in melting point temperatures, 1209K for germanium and 1687K for silicon. Beside the phase boundary at the surface, the heat loss of germanium is reduced by a factor of four considering Stefan-Boltzmann's law for the melting point temperatures. Accordingly, the temperature gradient for germanium is lower and the crystallization phase boundary is more sensitive to temperature fluctuations. Concordantly heat loss from the melt is lower than for the solid body, because of the lower emissivity of the liquid, therefore, deviations from a horizontal three phase boundary are stabilized or get even emphasized, which results in bulge growth. From this consideration the induction coil was chosen to be conical, to reduce the coupling of the coil with the melt and, therefore, reduce the heating of the rim. The effect was not strong enough and also an increase of the angle, for intensifying this effect did not lead to the expected result.

\begin{figure}[hbtp]
\centering{
\includegraphics[width=7cm]{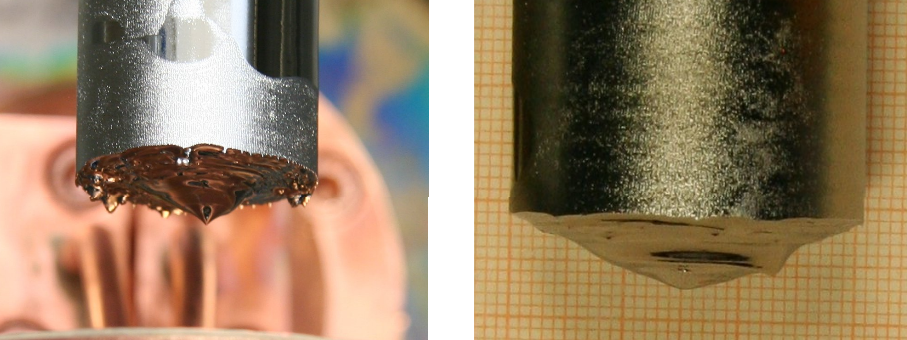}
}
\caption{The melting interface of a 30mm germanium feed rod under argon-gas (right) and under helium-gas (left) with spikes \label{fig:vorratsstaebe}}
\end{figure}

With the same goal of cooling the rim, argon gas was used for convective cooling by blowing 5-10l/min to the side of the crystal. The maximum volume flow and also an increase to four nozzles did not show the desired effect. Only a shaking of the melt surface was visible. Another mean to increase the cooling effect of the gas is the increase of the pressure. Unfortunately this evinced the problem of spike creation. Spikes (fig. \ref{fig:vorratsstaebe}a) are small unmolten parts of the feed rod, which are not inductively heated anymore. Consequently they touch the induction coil because of the downward movement, which breaks the process mostly by spilling out the melt. Anyhow, this was the first hint, that gas cooling has an influence.

\begin{figure}[hbtp]
\centering{
\includegraphics[width=7cm]{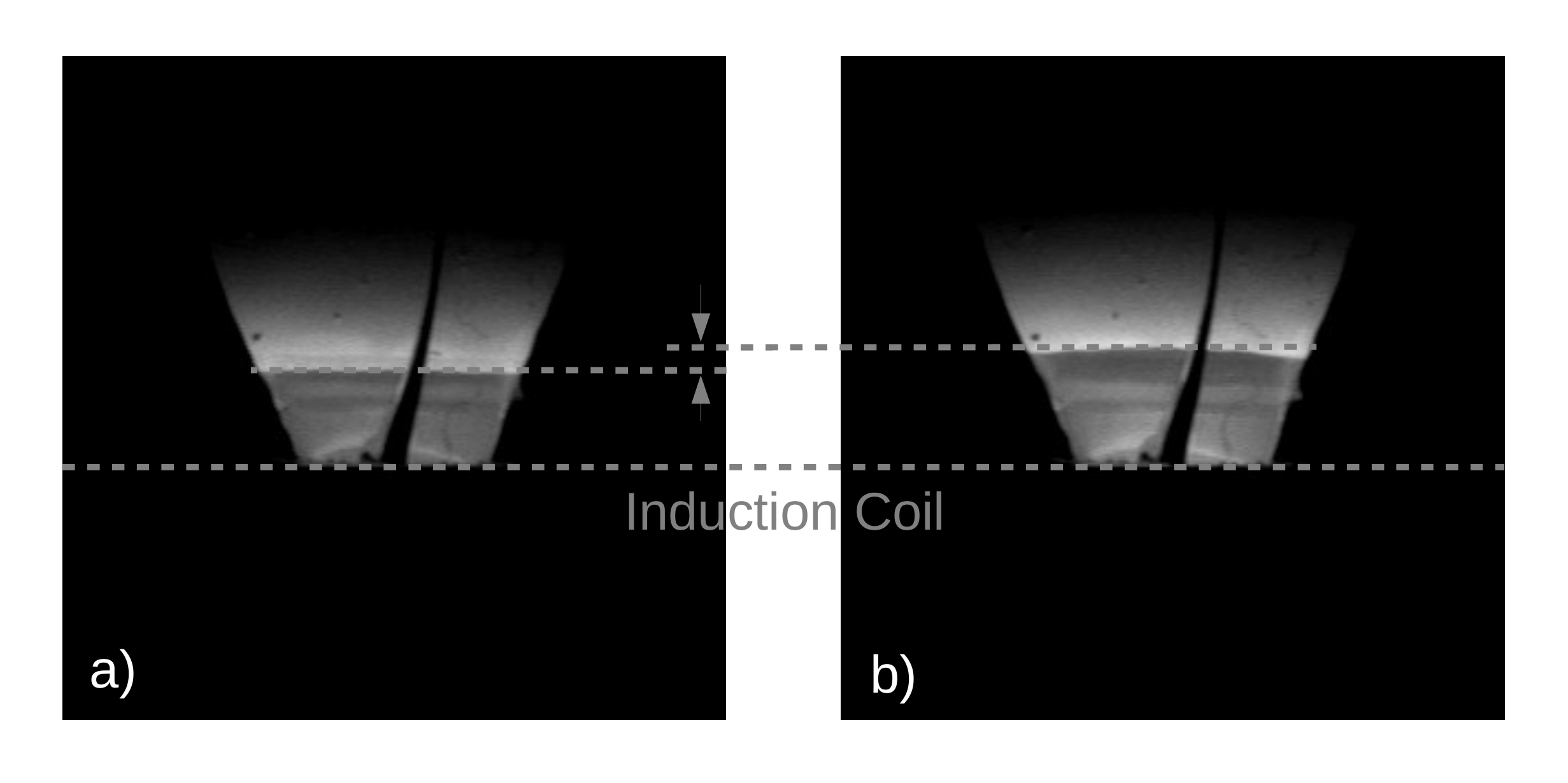}
}
\caption{Melt drop hanging from the feed rod in helium-gas a) and argon-gas b) under elsewise same conditions\label{fig:tropfen}}
\end{figure}

Firstly in helium-gas, the diameter of the crystal could be increased until 26mm without screws. The 10 times higher heat conductivity compared to argon increases the cooling and stabilizes the growth. Unfortunately the spikes also arose (fig. \ref{fig:vorratsstaebe}a), after melting the cone of the feed rod. So only a few millimeters could be grown with this diameter. The cooling effect could be also verified by observing a melt drop under different atmospheres. In the same experiment the atmosphere of the gas was exchanged and for the same position and heater power the melt drop is bigger in argon, because the heat loss is reduced and so, more material could be molten (fig. \ref{fig:tropfen}).

To get longer crystals we tried different methods to get rid of the spikes, which work successfully for silicon. But neither increasing the frequency of the generator from 2.6MHz to 3.6MHz nor an additional radiation heating with a round tube emitter nor a mixture of argon and helium solved the problem.

\begin{figure}[hbtp]
\centering{
\includegraphics[width=5cm]{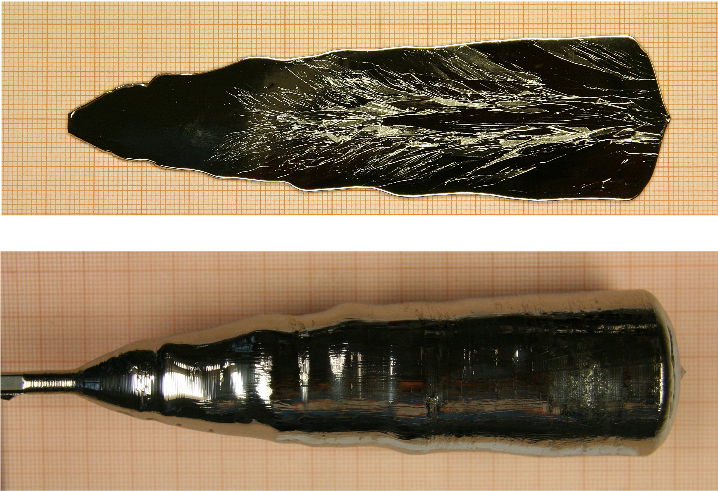}
}
\caption{Germanium crystal of 35mm in maximal diameter (down) with an axial-longitudinal cut after structural etching. After 20mm the material is not single crystalline anymore.\label{fig:fzkristall}}
\end{figure}

The breakthrough was achieved with the idea of separate gases around the feed rod, mainly argon, and the crystal, mainly helium for increased cooling. By this means it was concurrently possible to smoothly melt the source material and suppress the screw of the growing crystal. First-time, a stable and stationary germanium FZ process with a diameter of 35mm could be realized. Crystal diameter and length were limited by the feed rod diameter of 30mm and a length of several centimeters. Sadly, a single crystal-crystalline growth was only possible until a diameter of 20mm (see fig. \ref{fig:fzkristall}). It should be possible, as for silicon, to use the Dash method and grow a dislocation-free crystal after pulling a thin neck.

\vspace{-0.3cm}
\section{Conclusion}
The results for the growth of germanium crystals are promising for the use of crucible-free methods, despite of germanium's disadvantageous material properties for this process type. 

The FZ germanium process was only possible under quite different growth conditions. Only the simultaneous use of helium around the crystal and argon around the feed rod pointed out to be a solution for the instable growth and melting behavior. For the future the used setup will be scaled up to increase the crystal diameter. Simultaneously it will be investigated, if the Dash method will allow to get a single crystal similar to the silicon FZ process.

The pedestal method was more easily adaptable to the new material by slightly changing the induction coil. The numerical simulation gives some understanding of the process, which helps to optimize the process control. The experimental results were in good agreement with them as well. An option for a further increase in diameter would be to use helium for the crystal cooling as in the FZ case.

\vspace{-0.3cm}
\bibliographystyle{elsarticle-harv} 

\end{document}